\documentclass[prb,twocolumn,showkeys,showpacs,preprintnumbers,floatfix,english]{revtex4}
\usepackage[utf8]{inputenc}
\usepackage[T1]{fontenc}
\usepackage{babel}
\usepackage{ifpdf}
\usepackage{graphicx}
\usepackage{color}
\usepackage{lineno}
\usepackage{pslatex}
\usepackage{bm}
\usepackage{dcolumn}
\usepackage{amsmath}
\usepackage{amssymb}
\usepackage{amscd}
\ifpdf
\usepackage{epstopdf}
\usepackage[%
pdftitle={lGe2Se3},%
pdfauthor={},%
pdfsubject={lGe2Se3},%
pdfkeywords={lGe2Se3},%
pdfstartview=FitH,%
bookmarks=true,%
bookmarksopen=true,%
breaklinks=true,%
colorlinks=true,%
linkcolor=blue,anchorcolor=blue,%
citecolor=blue,filecolor=blue,%
menucolor=blue,pagecolor=blue,%
urlcolor=blue]{hyperref}
\else
\usepackage[%
pdftitle={lGe2Se3},%
pdfauthor={},%
pdfsubject={lGe2Se3},%
pdfkeywords={lGe3Se3},%
breaklinks=true,%
colorlinks=true,%
linkcolor=blue,anchorcolor=blue,%
citecolor=blue,filecolor=blue,%
menucolor=blue,pagecolor=blue,%
urlcolor=blue]{hyperref}
\fi
\include{bibsymb}
\usepackage{enumerate}
\usepackage{multirow}
\usepackage{tabls}
\newcommand{\ddst}{false}

\begin{document}
\draft
\title{Viscosity and viscosity anomalies of model silicates and magmas: a numerical investigation}
\author{M. Bauchy, B. Guillot, M. Micoulaut\footnote{Corresponding author: mmi@lptmc.jussieu.fr}, N. Sator}
\affiliation{
Laboratoire de Physique Th\'{e}orique de la Mati\`ere Condensée,
Universit\'{e} Pierre et Marie Curie, 4 Place Jussieu, F-75252 Paris Cedex 05, France
}
\date{\today}
\begin{abstract}
We present results for transport properties (diffusion and viscosity) using computer simulations. Focus is made on a densified binary sodium disilicate 2SiO$_2$-Na$_2$O (NS2) liquid and on multicomponent magmatic liquids (MORB, basalt). In the NS2 liquid, results show that a certain number of anomalies appear when the system is densified: the usual diffusivity maxima/minima is found for the network-forming ions (Si,O) whereas the sodium atom displays three distinct régimes for diffusion. Some of these features can be correlated with the obtained viscosity anomaly under pressure, the latter being be fairly well reproduced from the simulated diffusion constant. In model magmas (MORB liquid), we find a plateau followed by a continuous increase of the viscosity with pressure. Finally, having computed both diffusion and viscosity independently, we can discuss the validity of the Eyring equation for viscosity which relates diffusion and viscosity. It is shown that it can be considered as valid in melts with a high viscosity. On the overall, these results highlight the difficulty of establishing a firm relationship between dynamics, structure and thermodynamics in complex liquids. 
\end{abstract}
\maketitle
\section{Introduction}
Viscosity is one of the key properties influencing the overall behavior of magmatic liquids. It is a fundamental property in Earth Sciences and has therefore led after nearly five decades of intensive research to a huge body of experimental and theoretical studies and data base (Giordano et al., 2008; Mysen and Richet, 2005). Systematic investigations with composition and temperature in simple or complex silicates have been performed, and some generic trends have been identified that have become quite popular. For instance, it has been found that numerous systems were displaying an Arrhenius behavior (Micoulaut, 2010) with temperature of the form $\eta=\eta_0\exp[\text{E}_\text{A}/k_B\text{T}]$ where E$_\text{A}$ represents the activation energy for viscous flow. When properely rescaled with a reference temperature, T$_g$, at which the viscosity is 10$^{12}$ Poise, two categories of liquids have been identified (Angell, 1995) : those which remain Arrhenius-like in the super-cooled liquid over the entire range of temperatures implying that E$_\text{A}$ does not depend on T, and those which display a curvature in the Arrhenius representation (a semi-log plot as a function of inverse temperature) indicating that one has a temperature dependence in the activation energy for viscous flow (Wang et al., 2006). Usually, such liquids are fitted quite successfully with a Tamman-Vogel-Fulcher (TVF) law (Pathmanathan and Johari, 1990; Giordano and Dingwell, 2003a,b) : $\eta=\eta_0\exp(A/(\text{T}-\text{T}_0)$, $A$ and T$_0$ being constants.
\par
There has been a continuing effort to measure viscosity as function of pressure (Scarfe, 1973; Kushiro, 1976, 1978; 1986; Kushiro et al., 1976; Scarfe et al., 1979; Brearley et al., 1986; Dunn and Scarfe, 1986; Mori et al., 2000; Suzuki et al., 2002, 2005, 2011; Reid et al., 2001; 2003; Tinker et al., 2004; Liebske et al., 2005; Ardia et al., 2008; Del Gaudio and Behrens, 2009). At constant temperature, the evolution of viscosity with pressure is complex and depends on the composition of the silicate melt, in particular on its degree of depolymerization. With highly polymerized melts (e.g. albite and dacite) the viscosity decreases continuously with pressure, though its behavior above the maximal pressure of investigation ($\simeq$7~GPa for albite and dacite) cannot be clearly anticipated. However, it is observed that the decrease of the viscosity with pressure becomes more gradual when the temperature is raised and can lead to a plateau value at high pressure (e.g. jadeite). When the melt is less polymerized (e.g. albite-diopside system), after an initial decrease, and according to the temperature, the viscosity exhibits a plateau value or a weakly pronounced minimum with the pressure. With depolymerized melts (e.g. diopside and peridotite) a different situation occurs. The viscosity first increases with rising pressure and reaches a maximum value prior to decrease at higher pressure, but higher the degree of depolymerization of the melt more pronounced is the viscosity maximum. Moreover, in a more general way, the pressure evolution of the viscosity has been found to be (anti-) correlated with that of the oxygen and silicon diffusion coefficients (Shimizu and Kushiro, 1991; Poe et al., 1997; Reid et al., 2001; Tinker et al., 2003), a finding which has been interpreted as the signature of a pressure-induced structural rearrangement.
 \par
The understanding of changes in the viscous flow with applied pressure can be suitably investigated by molecular dynamics (MD) simulations. There are several ways to compute viscosity. Readers should refer to Allen and Tildesley (1987) for a detailed discussion and presentation. At equilibrium, the computation of viscosity from MD simulations can be performed by using the Green-Kubo (GK) formalism which is based on the calculation of the stress auto-correlation function (Boon and Yip, 1980) given by:
\begin{equation}
F(t)=\langle P_{\alpha\beta}(t) P_{\alpha\beta}(0) \rangle
\label{equf}
\end{equation}
where $P_{\alpha\beta}(t)$ is the $\alpha\beta$ component ($\alpha,\beta$)=(x,y,z) of the molecular stress tensor defined by:
\begin{equation}
P_{\alpha\beta}=\sum_{i=1}^{N}m_i v_i^{\alpha} v_i^{\beta} +
\sum_{i=1}^N \sum_{j>i}^N F_{ij}^{\alpha} r_{ij}^{\beta}\quad
\alpha \neq \beta,
\label{equ2}
\end{equation}
where F$_{ij}^{\alpha}$ is the component $\alpha$ of the force between
the ions $i$ and $j$, $r_{ij}^{\beta}$ and $v_i^\beta$ being the $\beta$ component of the distance between two atoms $i$ and $j$, and the velocity of atom $i$, respectively.
Alternative forms for $P_{\alpha\beta}$ exist (Allen and Tildesley, 1987) but these are found to lead to spurious effects when periodic boundary conditions are considered (Lee, 2007) or when the Ewald sum is not performed in a neat way (Alejandre et al., 1995; Allen et al., 1994). In this case, $F(t)$ is found to not decay to zero in the long-time limit, as expected (Lee, 2007; Green, 1957; Kubo, 1957). One other possibility is to compute the viscosity within the framework of Non-Equilibrium Molecular Dynamics (Lees and Edwards, 1972). Here, the ratio of shear stress to a strain rate is computed and extrapolated to the limit of zero driving force (Borodin et al., 2009; Fernandez et al. 2004; Cherne and Deymier, 1998).
\par 
Coming back to the GK formalism commonly used with MD simulations, the viscosity is calculated from the time integral of the stress auto-correlation function $F(t)$ (see eqs. (\ref{equf}) and (\ref{equ2})),
\begin{equation}
\eta=\frac{1}{k_BTV}\int_0^{\infty} F(t) dt.
\label{equh}
\end{equation}
\par
The viscosity of a few silicate melts has been evaluated by MD simulation in following this route. Ogawa et al. (1990) were the first to evaluate the viscosity of a silicate melt (a sodium disilicate) by MD simulation but their results are uncertain due to a poor statistics. For silica, Horbach and Kob (1999) determined the temperature behavior and Barrat et al. (1997) studied the pressure dependence at high temperatures but without pointing out explicitly that their results showed a viscosity minimum at about 20 GPa. More recently Lacks et al. (2007) in investigating by MD the properties of silicate liquids along the MgO-SiO$_2$ join found a viscosity minimum at about 20 GPa in pure silica, a viscosity minimum which disappears progressively with increasing MgO content. Since then, these results have been confirmed by the MD studies of Adjaoud et al. (2008, 2011) on Mg$_2$SiO$_4$ melts, those of Spera et al. (2011) on molten enstatite and also by first-principles MD calculations of Karki and Stixrude (2010a,b) on liquid silica and on liquid enstatite. We finally mention the recent studies of Karki et al. (2011) and Gosh and Karki (2011) on anorthite liquid and Mg$_2$SiO$_4$ liquid, respectively.

\section{Modeling silicates}
\subsection{Simulation details}
To complement the above MD studies and to bring some new information, we have investigated in the present study two silicate melts exhibiting comparable NBO/T ratio ($\simeq$ 0.8 and 1) but with very different chemical compositions; a sodium disilicate (NS2) and a mid-ocean ridge basalt (MORB). The molecular dynamics (MD) simulations were performed with the DL\_POLY 2.0 code (Smith and Forrester, 1996) in NPT Ensemble. The equations of motions for atoms were solved with a time step of 1-2 fs (10$^{-15}$~s) by the leap-frog algorithm. 
\par
Liquid Na$_2$O-2SiO$_2$ has been simulated by placing 666 silicon, 666 sodium, and 1668 oxygen atoms in a cubic box with periodic boundary conditions. As the atoms are bearing charges (ions), the long range coulombic forces are evaluated by a Ewald sum. The atoms interact via a two-body potential (Born-Majer type) parametrized by 
Teter (2003, see also Cormack et al., 2003), which writes
\begin{eqnarray}
\label{teter}
V_{ij}(r)=A_{ij}.\exp(-r/\rho_{ij})-C_{ij}/r^6 + z_{i} z_{j}/r,
\end{eqnarray}
where $r$ is the distance between atoms $i$ and $j$, $z_{i}$ is the effective charge associated with the ion $i$, and where $A_{ij}$, $\rho_{ij}$ and $C_{ij}$ are parameters describing repulsive and dispersive forces between the ions $i$ and $j$. In the present report, we do not focus on structure and thermodynamic properties of liquid NS2 simulated with this potential and for a detailed analysis and comparison with experimental data we refer the reader to Du and Cormack (2004, 2005). Notice that at zero pressure and 300K the density of the simulated NS2 glass is equal to 2.45~g/cm$^3$, i.e. quite close to the experimental value (Vaills et al., 2001) 2.37 g/cm$^3$.\par
\par
The mid-ocean ridge basalt (MORB), a nine component system (0.13 wt\% K$_2$O, 2.94 wt\% Na$_2$O, 11.87 wt\% CaO, 7.77 wt\% MgO, 8.39 wt\% FeO, 1.15 wt\% Fe$_2$O$_3$, 15.11 wt\% Al$_2$O$_3$, 1.52 wt\% TiO$_2$ and 50.59 wt\% SiO$_2$), has been simulated with a two-body potential of the same form as eq. (\ref{teter}). The chemical composition of the MORB melt and the potential parameters are those given by Guillot and Sator (2007a) in their simulation study of natural silicate melts (see Tables 1 and 2 therein). The simulated sample of MORB was composed of 3,000 atoms (555 Si, 12 Ti, 195 Al, 9 Fe$^{3+}$, 60 Fe$^{2+}$, 126 Mg, 141 Ca, 63 Na, 3 K, and 1,836 O) contained in a cubic box with periodic boundary conditions. The control parameters (time step, Ensemble, etc.) of the simulation are the same as for liquid NS2. The density of MORB at liquidus temperature and its equation of state (i.e. density versus pressure) are well reproduced by the MD calculation (for a detailed description of the properties of the MORB melt, see Guillot and Sator 2007a,b). Notice that long simulation runs ($\simeq$10~ns or 10$^7$ time steps) were performed to reach a good statistics when evaluating the viscosity by the Green-Kubo integrand.
\par
Before presenting results, it is worthwhile to keep in mind some key data when one is about to calculate the viscosity of a supercooled melt. In using a high performance computer, the longest MD run that can be performed (in a reasonable time) for a system composed of $\simeq$ 3,000 atoms is about 1 $\mu$~s (or 10$^9$ MD steps). To observe on this timescale the diffusion of atoms in the simulation box, the mean square displacement of each atom after 1 $\mu$~s has to be (on average) of the order of $\simeq$10~\AA$^2$ (the mean distance between two nearest neighbors is roughly 3~\AA\ in a melt, so the diffusive regime is reached when each atom has interchanged its position with a nearest neighbor). Therefore, the smallest diffusion coefficient that can be evaluated over a 1 $\mu$~s MD run is equal to $\simeq$ 2$\times$10$^{-14}$~m$^2$/s (from Einstein relation $D_{min}$=$\langle r^2_{min}\rangle/6t_{max}$ where $\langle r^2_{min}\rangle$=10~\AA$^2$ and t$_{max}$ = 10$^{-6}$~s). In applying the Eyring theory to viscous flow , the viscosity $\eta$ can be deduced from the diffusion coefficient $D$ of network forming ions through the equation,
\begin{eqnarray}
\label{eyring}
\eta={\frac {k_BT}{\lambda D}}
\end{eqnarray}
where T is the temperature and $\lambda$ a jump distance in the viscous flow. Experimentally it has been shown (Shimizu and Kushiro, 1984) that this equation works well with viscous liquids (e.g. silicates with high silica contents) provided that the D value is assigned to that of O or Si atoms and $\lambda\simeq$2.8\AA, the oxygen-oxygen mean distance in silicate melts. Using equ. (\ref{eyring}), a value of D$_O$ equal to 2$\times$10$^{-14}$~m$^2$/s leads to a corresponding viscosity af about 3000~Pa.s at 1473~K. We can use the Maxwell relation for which the relaxation time for viscous flow is related to the viscosity through the equation, $\tau_{relax} = \eta/G_{\infty}$, where $G_{\infty}\simeq10^{10}$~Pa.s is the shear modulus at infinite frequency (Dingwell and Webb, 1989). So to fully relax a viscous melt of viscosity $\eta\simeq10^4$~Pa.s one needs to perform a MD run longer than $\tau_{relax}=1 \mu s$, a finding consistent with the evaluation based on the diffusion. Although qualitative, these estimates are useful because they point out that only the low viscosity regime (i.e. $\eta$ $\ll$ 10$^4$ Pa.s) can be investigated by MD simulation, the highly viscous regime (10$^4$ - 10$^{12}$~Pa.s) being unreachable with usual numerical resources. In the present study our MD runs did not exceed $\simeq$10 ns, which means that only a viscosity value smaller than $\simeq$10~Pa.s (100 Poise) could be investigated (notice that most of the MD studies published in the literature often do not exceed 1 ns run intervals).
\par
\subsection{Diffusion coefficients }
We first compute the mean-square displacement of a tagged atom of type $\alpha$ in the melt, given by
\begin{equation}
\langle r^2(t) \rangle =\frac{1}{N_{\alpha}} \sum_{i=1}^{N_{\alpha}}
\langle |{\bf r}_i(t)-{\bf r}_i(0)|^2\rangle\quad,
\label{eqdiff}
\end{equation}
\begin{figure}
\includegraphics*[width=0.9\linewidth, keepaspectratio=true, draft=\ddst]{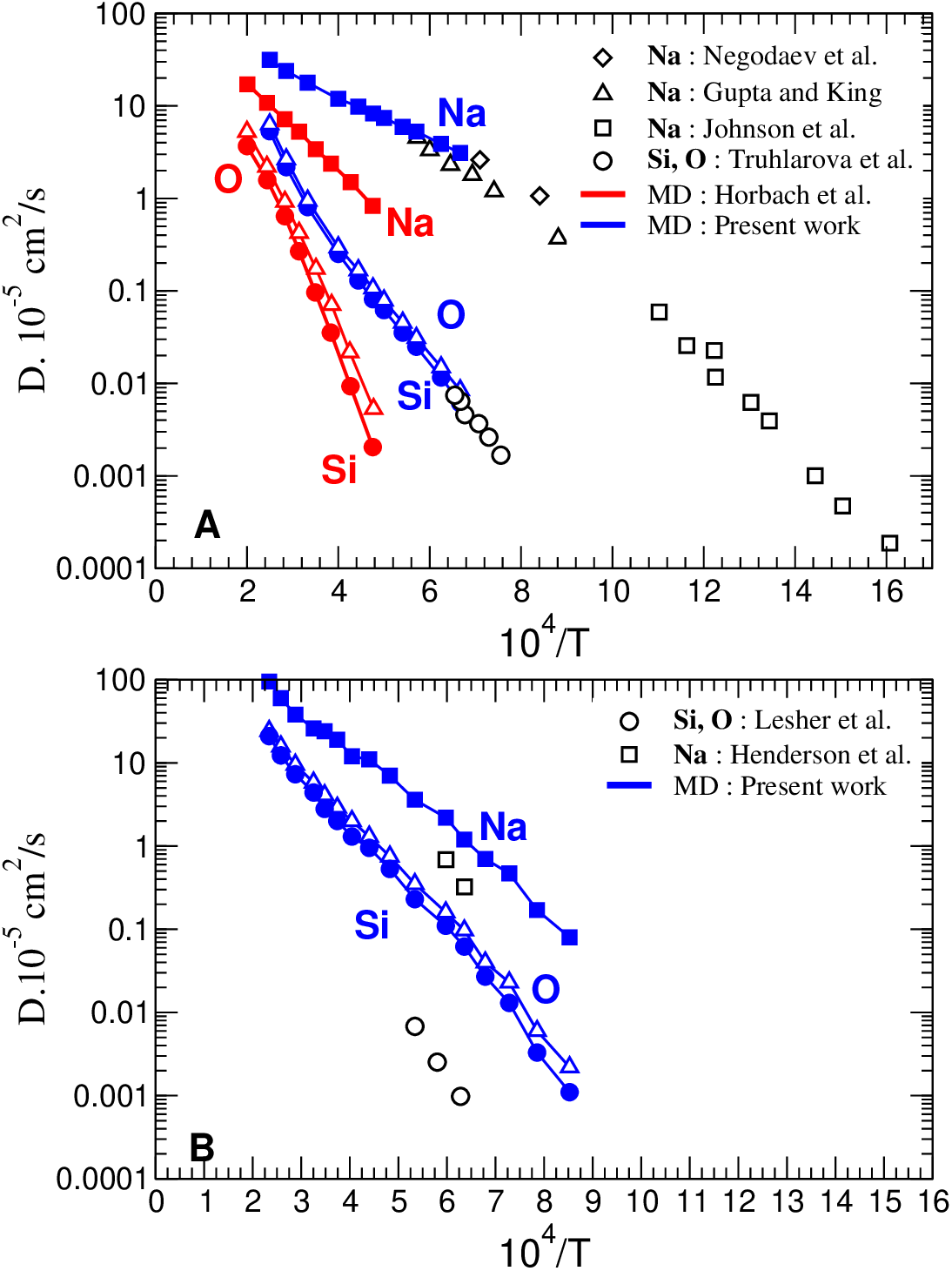}
\caption{\label{diff_compar} (Color online) A) Computed diffusion constants D$_{Na}$, D$_{Si}$ and $D_{O}$ in the NS2 liquid as a function of inverse temperature (blue curves and symbols), compared to experimental data 
for D$_{Na}$ (Gupta and King, 1967; Negodaev et al. 1972; Johnson et al., 1951) and D$_{Si}$, D$_O$ (Truhlarova et al., 1970), and to the simulated values of D$_{Na}$, D$_{Si}$ and D$_O$ using an alternative potential (red curves and symbols, Horbach et al., 2001). B) Computed diffusion constants D$_{Na}$, D$_{Si}$ and $D_{O}$ in the MORB liquid as a function of inverse temperature (blue curves and symbols), compared to experimental data for D$_{Na}$ (Henderson et al.) and D$_{Si}$, D$_O$ (Lesher et al. 1996).
}
\end{figure}
and extract from the dependence of $\langle r^2(t) \rangle$ the long time 
behavior where the dynamics becomes diffusive. Using the Einstein relation 
$\lim_{t\to \infty} \langle r^2(t)\rangle/6t=\text{D}$, one can indeed have access to the diffusion 
constants D$_i$ ($i$=Si, O, Na) from the mean square displacement. These quantities are plotted for NS2 and MORB in Fig. \ref{diff_compar} as a function of inverse temperature. 
Because of the slowing down of the dynamics at low temperature, the computation of D$_i$ is here restricted to 
T $>$ 1500~K with NS2 and T $>$ 1173~K with MORB. For NS2, the diffusion of sodium atoms is found to be remarkably close to the experimental data obtained by Gupta and King (1967) and Negodaev et al. (1972), with a very good agreement somewhere around $10^4/T\simeq 6$. For network forming ions (Si,O) an excellent agreement is found with the experimental data of Truhlarova et al. (1970) who determined the (Si,O) diffusion from a SiO$_2$ dissolvation reaction in a NS2 liquid. In this respect, the Teter potential appears to be very accurate for reproducing the diffusion of elements in the NS2 melt. We remark that the diffusion coefficients display an Arrhenius-like dependence when 10$^4$/T$>$4 with activation energies equal to 1.18~eV, 1.17~eV and 0.43~eV for Si, O and Na respectively.
\par
To be complete, one should stress that alternative potentials (Kramer et al., 1991) lead to diffusion coefficients for (Na,Si,O) that are found (Horbach et al., 2001) to be at least one order of magnitude lower than the experimental values (for 
10$^4$/T$>$5), and our own results on the MORB liquid as well (see below), whereas the corresponding activation energies also differ by a factor of $\simeq$2 (Fig. \ref{diff_compar}). These findings illustrate how difficult the numerical reproduction of diffusion coefficients in silicate melts can be. This well-known feature has been highlighted for the case of silica by Hemmati and Angell (1997) who showed that diffusion is highly model dependent with differences between models increasing up to at least three orders of magnitude when the liquid is cooled down. So, the very good agreement between experimental and simulated ionic diffusivities in the NS2 melt, allows one to be reasonably confident in the ability of the Teter potential to describe accurately other transport properties as the melt viscosity.
As for the MORB melt, the temperature dependence of the O, Si and Na diffusion coefficients is shown in panel B of Fig. \ref{diff_compar} (among all the network modifying cations in MORB, only the diffusion coefficient of Na is shown because it is the most mobile ion and because it is common with the NS2 melt). Na is found to diffuse more slowly in MORB than in NS2 whereas Si and O atoms are diffusing faster. However, the agreement between simulated and experimental values is poorer with MORB than with NS2 (see Fig. \ref{diff_compar} for a comparison). Actually, Na, Si and O atoms diffuse too rapidly in the simulated MORB (by a factor of $\simeq$5 for Na and $\simeq$50 for Si and O ). So one may anticipate some deviation between calculated and experimental viscosities for the basaltic liquid, and a better agreement for liquid NS2.
\subsection{Viscosity }
\begin{figure}
\includegraphics*[width=0.9\linewidth, keepaspectratio=true, draft=\ddst]{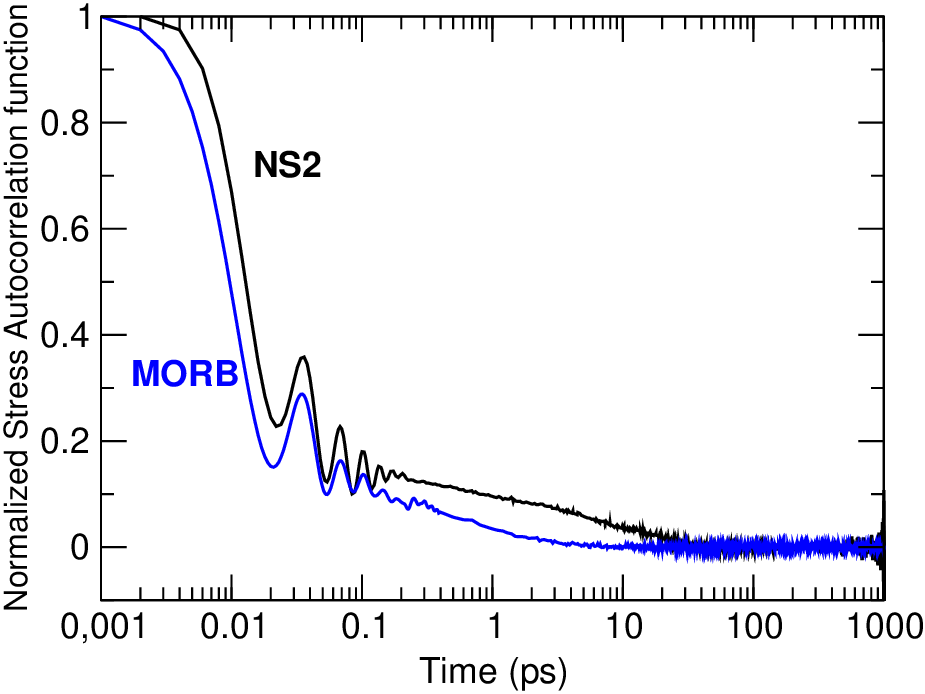}
\caption{\label{stress} (Color online) Stress auto-correlation function $F(t)$ of a 2000 K liquid NS2 as a function of time for $\rho$=2.5 g/cm$^3$, and a 2273 K MORB liquid at the same density.
}
\end{figure}
\begin{figure}
\includegraphics*[width=0.9\linewidth, keepaspectratio=true, draft=\ddst]{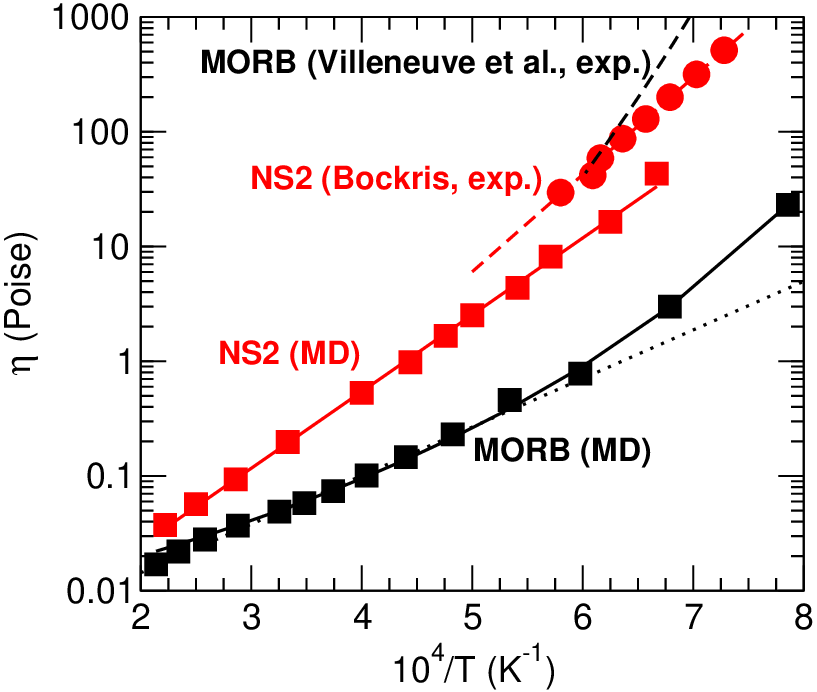}
\caption{\label{visco_T} (Color online) Simulated viscosity of the NS2 liquid at zero pressure (filled red squares), compared to experimental data (red circles) from Bockris et al. (1955), together with simulated viscosity of the MORB liquid at zero pressure (filled black squares), compared to experimental data (broken black curve) from Villeneuve et al. (2008). The dotted line is a high temperature Arrhenius fit for the MORB data whereas the solid line is a TVF fit (see text for parameters).
}
\end{figure}

In Figure \ref{stress}, we represent the stress auto-correlation function $F(t)$ of NS2 and MORB at around zero pressure, computed following equs. (\ref{equf}) and (\ref{equ2}). As mentioned above, the stress auto-correlation function is indeed found to decay to zero. Typical oscillations which are found at the subpicosecond timescale are associated with frequencies in the Terahertz range and above [10-300 THz] corresponding to the main band found in the vibrational density of states. This band is attributed to stretching motions of Si-O and Na-O bonds (and other cation-oxygen bonds in MORB). Note that these oscillations are damped as the pressure is increased (not shown). Following eq. (\ref{equh}), the time integration of the stress auto-correlation function leads to the viscosity (Fig. \ref{visco_T}). The viscosity of NS2 exhibits an Arrhenius-like behavior over a broad temperature range (2$<$10$^4$/T$<$6). The calculated activation energy for viscous flow is about $\simeq$1.33~eV which is in a reasonable agreement with Bockris data ($\simeq$1.65 eV) measured (Bockris et al., 1954) in the temperature range (1723-1373 K or 5.8$<$10$^4$/T$<$7.3). Furthermore, in the temperature range where calculated and experimental values can be compared with each other (i.e. 5.8$<$10$^4$/T$<$6.7) calculated values are smaller by a factor of $\simeq$2 than the experimental data of Bockris et al. (1955). This deviation is surpring given the good agreement obtained for the diffusion coefficients (see Figure \ref{diff_compar}), and given the fact that $\eta$ and $D$ can be simply related (equ. \ref{eyring}). 
\par
In the case of MORB the temperature dependence of the calculated viscosity is found to be non Arrhenian and can be fitted accurately over a large temperature range by the TVF equation with $\eta_{\infty}$=6.5$\times$~10$^{-3}$ Poise, $A$ = 0.42~eV, and $T_0$ = 670~K. Experimentally, the viscosity of basaltic melts is known to be non Arrhenian, so the simulation reproduces this feature, but the simulated MORB is much less viscous than the real one (a factor of 50 at 1673 K, see Fig. \ref{visco_T}), this is why $A$ and $T_0$ values are slightly different from those recommended for basalts by the regression formula of Giordano and Dingwell (2003a,b), and Giordano et al. (2008). In fact the MORB melt simulated at 1273 K corresponds, as far as the viscosity is concerned, to the real melt at about 1673K, this temperature deviation diminishing at higher temperatures where the viscosity of any silicate melt (whatever is its composition) tends to be very low (less than 0.1 Poise above 3000 K). More generally, the agreement between experimental and calculated viscosities for silicate melts is actually difficult to obtain because the viscosity seems to be even more sensitive to the details of the potential than the diffusion coefficients themselves (for a related discussion see Vuilleumier et al., 2009). Moreover, direct comparisons between experimental and simulated values of the viscosity are scarce in the literature (e.g. Lacks et al., 2007; Adjaoud et al., 2008; Karki and Stixrude, 2010b, Spera et al., 2011) and are usually done at very high temperature ($>$2500K) where the viscosity is very low and its experimental value badly known or obtained from uncontrolled high-temperature extrapolations. 
\par
In the TVF equation, $\eta_{\infty}$ represents the high temperature limit to silicate melt viscosity. By analyzing measured viscosity curves for a great number of silicate liquids, it has been shown  that all silicate melts converge to a common value of viscosity at very high-T, about 10$^{-2}$-10$^{-3}$ Poise (Russell et al., 2003; Giordano et al., 2008, Zheng et al., 2011). But it is difficult to test this conjecture because of a lack of viscosity data for silicate liquids at super-liquidus temperatures (i.e. for T $\geq$ 2000 K). In contrast, the simulation can shed some light on this topic because the high-T range is easily accessible in a numerical experiment. However it is worthwhile to clarify what means the concept of a high-temperature limit for a silicate liquid. When a silicate melt is heated isobarically (e.g. under atmospheric pressure) it vaporizes incongruently in forming a gaseous phase which is a complex mixture of atoms and molecules (e.g. the major vapor species above liquid silica are SiO, SiO$_2$, O$_2$ and O, and about thirty species compose the vapor phase above a basaltic lava, for a review see Schaefer and Fegley, 2004). Nevertheless this is not a real problem as long as the resulting vapor pressure remains low, i.e. the temperature is not too high (e.g. the vapor pressure above basaltic lavas is 10$^{-5}$-10$^{-4}$ bar at 1900 K). This situation changes drastically at higher temperature where the vapor pressure becomes high (e.g. $\simeq$10$^{-2}$ bar at 2500K above for a tholeiitic liquid). Then the vaporized fraction  becomes rapidly non-negligible (10\% or more) and the composition of the residual melt deviates significantly from the original one. To avoid this problem, one only has to prevent vapor formation in applying on the liquid an isostatic pressure larger than the saturating vapor pressure. A convenient thermodynamic path consists to follow (or to be as close as possible) the liquid branch of the liquid-vapor coexistence curve up to the critical point. The critical point parameters for silicates have never been measured because of the very high temperatures involved (much beyond 5000 K, estimations for SiO$_2$ are given in Guissani and Guillot (1994) and Melosh (2007)). By imposing the pressure, we have been able to evaluate the viscosity of our simulated MORB along (or close by) the liquid branch of the liquid-vapor coexistence curve. However due to the density fluctuations generated by the MD calculation it becomes more and more difficult to preserve the cohesion of the liquid phase in rising the temperature because the liquid then tends to transform spontaneously into the corresponding vapor phase. So the highest temperature reached in keeping the cohesion of the liquid phase around 1 bar was 4673K, a temperature at which the density is equal to 1.47~g/cm$^3$ and the viscosity is evaluated to 10$^{-2.8}$ Poise. This latter value matches rather well the estimated range for $\eta_{\infty}$ (10$^{-2}$-10$^{-3}$ Poise, see above). It is certainly possible to pursue the calculation at higher temperature up to the critical point, but it is a tedious task to locate accurately the critical point of our model and we leave that to a future work. In using the same procedure with liquid NS2 we have reached 4500K, a temperature at which the liquid density is as low as 1.66 g/cm$^3$ and its viscosity 10$^{-2.5}$ Poise. We anticipate that the common value of $\eta_{\infty}$ for MORB and NS2 at the critical point could be about 10$^{-3}$ Poise or slightly less. Notice that in approaching the critical point, the structure of a silicate liquid is quite different of the structure from the melt near the liquidus. The near critical liquid is indeed fully depolymerized and is composed of molecular clusters involving various species (oxides). So it is expected that all silicate liquids will exhibit similar flow behavior in the critical region, that of a regular molecular liquid.

\par
\begin{figure}
\includegraphics*[width=\linewidth, keepaspectratio=true, draft=\ddst]{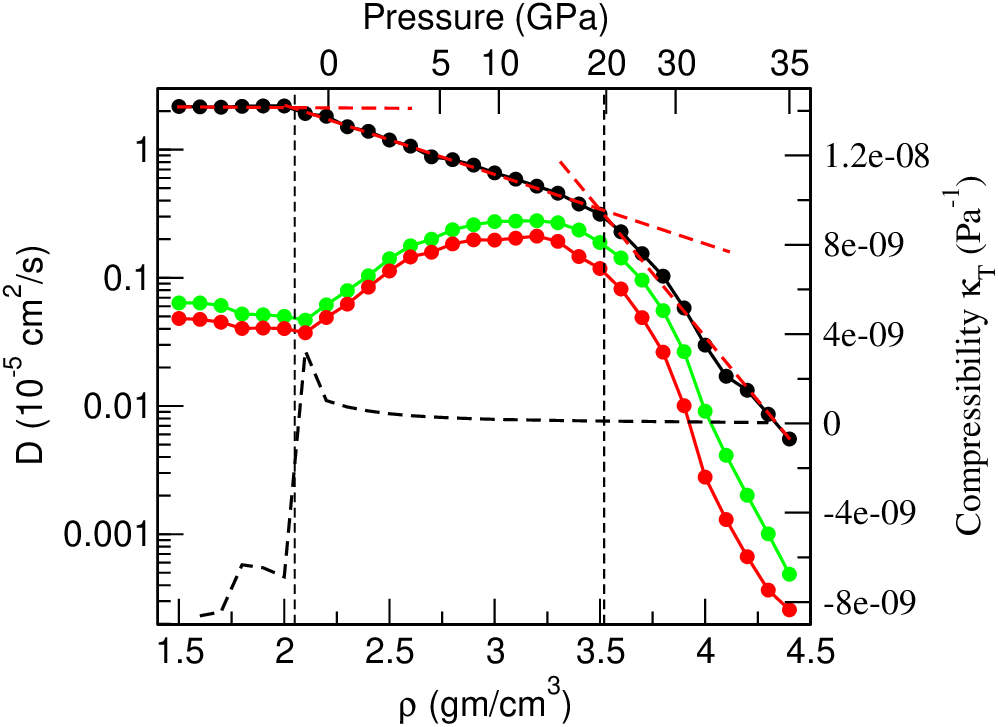}
\caption{\label{diffp} (Color online) Silicon (red), Oxygen (green), and sodium (black) diffusion constants of a NS2 liquid at T=2000 K as a function of system density (bottom axis) and pressure (top axis). Right axis: Isothermal compressibility of the liquid computed from the equation of state at 2000~K.
}
\end{figure}
\par
\begin{figure}
\includegraphics*[width=0.8\linewidth, keepaspectratio=true, draft=\ddst]{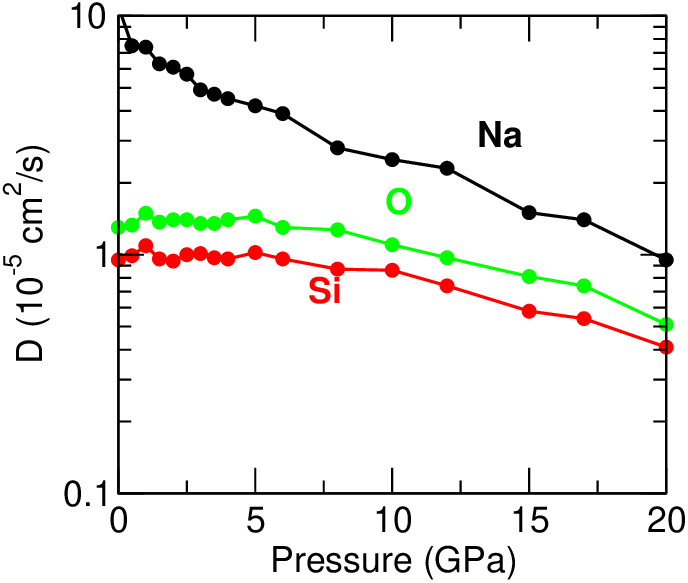}
\caption{\label{diffmorb} (Color online) Silicon (red), Oxygen (green), and sodium (black) diffusion constants of a MORB liquid at T=2273 K as a function of pressure.
}
\end{figure}

We now follow both viscosity and diffusion with pressure in our two melts (NS2 and MORB). In Fig. \ref{diffp} and \ref{diffmorb}, we plot the behavior of the diffusion constants with density and pressure. The most mobile atom, Na, is found to display the same behavior in the two melts: its diffusion coefficient is continuously decreasing with the pressure. In the case of MORB the other network modifying cations (Ca, K, Mg,...) show the same behavior with the pressure as Na does (Fig. \ref{diffmorb}).
\par
The behavior of network forming ions (Si,O) is quite different as their diffusion coefficients go through a maximum value with the pressure. Furthermore this maximum is located at about 15~GPa in NS2 and about 50~GPa in MORB and it is much more pronounced in the NS2 melt than in the basaltic composition. Notice that a diffusivity maximum for network forming ions is also observed with silica (Shell et al. 2002), germania (Dreyfus and Micoulaut, 2012), some sodo-aluminosilicates (Poe et al., 1997) and in water (Errington and Debenedetti, 2001). 
\par 
\begin{figure}
\includegraphics*[width=\linewidth, keepaspectratio=true, draft=\ddst]{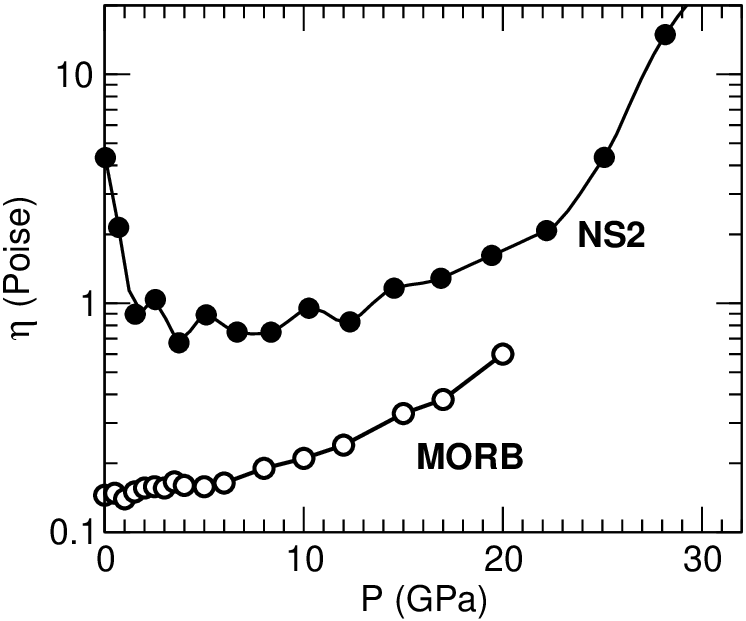}
\caption{\label{viscop} (Color online) Simulated viscosity of a NS2 liquid at T=2000 K (filled circles) and a MORB liquid at T=2273~K (open circles) as a function of pressure.
}
\end{figure}
Is such an anomaly also visible in the evolution of viscosity with pressure ?
Actually, for NS2 we do find that $\eta$ displays a minimum of viscosity but in a pressure range 4-10~GPa which does not coincide with the diffusivity maximum for Si and O (P$_{max}$=15~GPa). For MORB, the viscosity shows no minimum, it is nearly constant up to 5~GPa and increases progressively at a higher pressure. These simple findings suggest that the diffusivities of network forming ions are not the only ones responsible of the viscous flow, network modifying cations play also a role.

\section{Discussion}

As briefly recalled in the introduction, the pressure dependence of the viscosity has been reported experimentally for a number of silicates (Scarfe, 1973; Kushiro, 1976, 1978; 1986; Kushiro et al., 1976; Brearley et al., 1986; Dunn and Scarfe, 1986; Mori et al., 2000; Suzuki et al., 2002, 2005, 2011; Reid et al., 2003; Tinker et al., 2004; Liebske et al., 2005; Ardia et al., 2008). The analysis of these data along the sequence of composition albite$\rightarrow$ jadeite$\rightarrow$ rhyolite$\rightarrow$ andesite$\rightarrow$ basalts$\rightarrow$ albite$_x$-diopside$_{1-x}$ $\rightarrow$ jadeite$_x$-diopside$_{1-x}$ $\rightarrow$ diopside $\rightarrow$ peridotite, which covers a large NBO/T range (from 0 to 2.8), leads to a contrasted picture. In the initial pressure range (0-2~GPa), the slope (d$\eta$/dP)$_T$ is negative for composition with NBO/T $\leq$ 1.2 and is positive for higher NBO/T value. Although the NBO/T ratio is a good indicator of melt polymerization, its use for predicting the pressure evolution of viscosity requires some caution (e.g. Toplis and Dingwell, 2004). For instance, albite and jadeite melts are both considered as fully polymerized (NBO/T=0) but the pressure behavior of their viscosity is different (a viscosity curve presenting a weak minimum located about 2-3~GPa for jadeite as compared with a continuous decrease of the viscosity over several decades up to 7~GPa for albite, see Suzuki et al. 2002, 2011). In keeping in mind this information, and according to the above data analysis, the viscosity of NS2 and MORB composition (NBO/T=1 and 0.8) should exhibit a negative slope. For instance, Scarfe et al. (1987) notice that the viscosity of the NS2 melt is decreasing between 0 and 2~GPa. In the same way, the viscosity of oceanic basalts (Kushiro et al., 1976; Kushiro, 1986; Ando, 2003) maintained at fixed temperature decreases slightly with increasing pressure up to 3~GPa where it reaches a minimum value prior increase upon further compression. But as an increase of the temperature tends to decrease the viscosity of a melt maintained at a fixed pressure, the weak minimum of viscosity in basalts is expected to disappear in increasing the temperature. This is why a viscosity minimum is barely visible with our MORB melt simulated at 2273K, especially as the latter one describes the viscous flow of a real MORB at a higher temperature, may be as high as 2600-2700 K (see Fig. \ref{visco_T}). In contrast the viscosity of the simulated NS2 melt at 2000 K presents a clear minimum as a function of pressure (notice that the temperature dependence of the viscosity of the NS2 melt is much better reproduced by the simulation than in the MORB case, see Fig. \ref{visco_T}). 
\par
Since the first report of an anomalous pressure dependence of the viscosity ((d$\eta$/dP)$_T$ $<$0) by Kushiro (1976) and Kushiro et al. (1976), different authors have tried to explain the origin of this phenomenon. Using the Adam-Gibbs theory and a pressure dependence for the degree of melt polymerization, Bottinga and Richet (1995) have shown that the decrease of the specific volume with alkali content was leading to an increased sensitivity of the viscosity with pressure change. Gupta (1987) has proposed that a system having a thermal expansion coefficient in a glass larger than that of a liquid could display such anomalies. However, this does not seem to be verified experimentally (Knoche et al., 1994). A relationship with structural changes under pressure also has been proposed, in particular a pressure-induced coordination change of Si and Al or Si-O and Al-O bond weakening due to pressure-induced distortion of the liquid structure (Waff, 1975; Mysen et al., 1980). However, a study (Sharma et al., 1979) on dense liquid GeO$_2$ where both coordination change (Micoulaut et al. 2006) and a pressure induced viscosity anomaly occur could not establish the correlation. This does not rule out the possibility of a structural origin for the anomaly. For instance, Suzuki et al. (2002) do find a correlation between the gradual decrease of the viscosity of albite with pressure up to 7~GPa and the increasing concentration of 5-coordinated Al. 
\par
The variation of $\eta$ along an isotherm also suggests that the corresponding activation energy for viscous flow E${A}$(P,T) is minimum. Here, for selected pressures in the NS2 system, we have computed $\eta$ with temperature and find that the activation for viscous flow should minimize indeed in the pressure window, i.e. we obtain E$_A$=1.33, 1.09 and 1.34~eV for P=0, 6, and 22 GPa, respectively. The link between E$_A$ minima and the presence of a stress-free state in corresponding low-temparature glassy networks has been established recently (Micoulaut, 2010), indicative of a flexible to rigid transition (Micoulaut and Phillips, 2003). Since it is known that at ambient pressure the NS2 is flexible (Vaills et al., 2005), one may obtain a pressure-induced flexible to rigid transition with increasing P (Trachenko et al. 2004), the minimum in E$_A$ (and $\eta$) being one of the expected salient features.
\par
In summary, all these studies point out the subtle interplay between structure, dynamics and thermodynamics of the silicate melt that leads to anomalous behavior in transport properties. 

\subsection{Viscosity versus diffusion }

\begin{figure}
\includegraphics*[width=\linewidth, keepaspectratio=true, draft=\ddst]{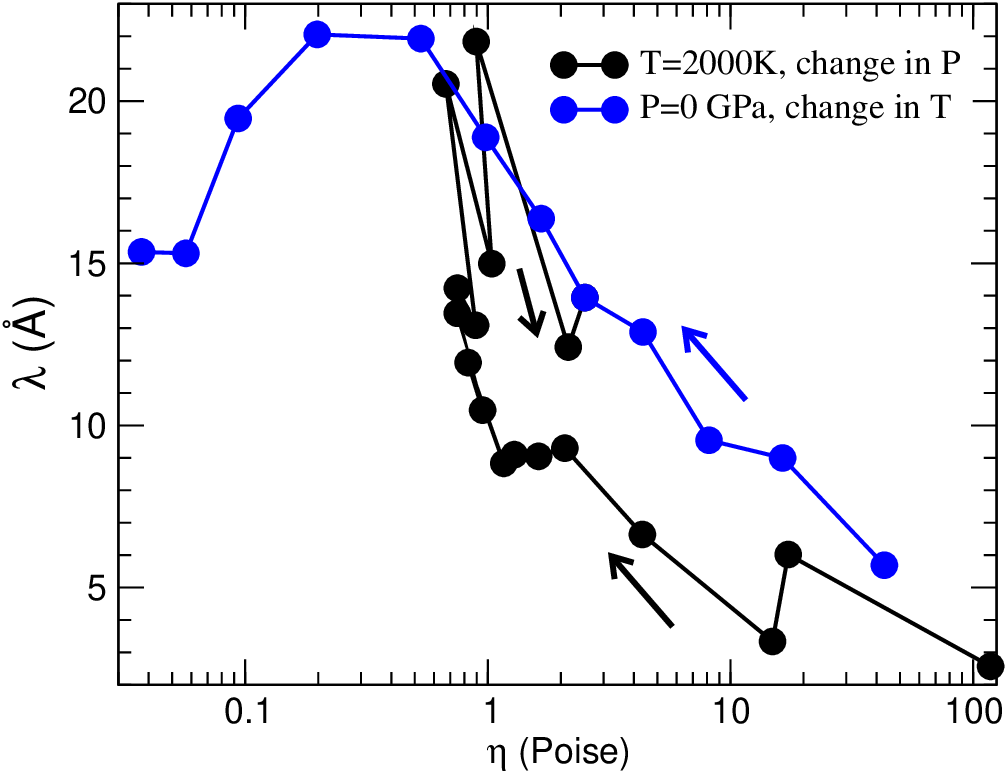}
\caption{\label{eyring_loglog} (Color online) Typical lengthscale $\lambda$ associated with the Eyring relation (\ref{eyring}) as a function of the melt viscosity of a NS2 liquid, computed either from the pressure variation (at 2000 K, black symbols) or from the temperature variation (at P=0, blue symbols).
}
\end{figure}

The connection between diffusion and viscosity can be realized in a simple way {\em via} the Eyring relationship given by eq. (\ref{eyring}) involving  $\lambda$ which is a typical jump
 distance for the diffusing atom. According to Eyring, the relationship (\ref{eyring}) holds
(i.e. $\lambda$ is constant and equal to the average interatomic distance) if the activated
process for diffusion can be assumed identical with that of viscous flow. So for silicate melts the network forming ions (Si and O) are good candidates to fulfill this prerequisite. In fact the Eyring relationship has been tested over a large range of melt composition, from simple binary oxides to natural magma compositions (Oishi et al., 1975; Yinnon and Cooper, 1980; Dunn, 1982; Shimizu and Kushiro, 1984; Dunn and Scarfe, 1986; Lesher et al., 1996; Reid et al., 2003; Tinker et al., 2004). In polymerized melts of high viscosity (e.g. jadeite), it is observed that the diffusivity of oxygen ions and the melt viscosity are related to each other through the Eyring relation in introducing for the jump distance, $\lambda$, the average oxygen-oxygen distance in the melt ($\simeq$~2.8\AA). In depolymerized melts where the viscosity is generally lower (e.g. basalts and diopside), the Eyring relation is fulfilled with $\lambda$ larger than 2-3 times the mean oxygen-oxygen distance. However even in polymerized melts, the Eyring relation fails to fully reproduce the correlated evolution of viscosity and oxygen self-diffusion coefficient. As a matter of fact it has been reported that oxygen self-diffusion coefficient in albite (Poe et al., 1997) and in dacite melts (Tinker and Lesher, 2001) reaches a maximum value at about 5~GPa whereas the viscosity of the corresponding liquids is continuously decreasing with the pressure and shows no viscosity minimum at 5~GPa (Suzuki et al., 2002; Tinker et al., 2004). In this context, we have evaluated the length, ${\lambda}=k_BT/ \eta D_{O,Si}$ along two thermodynamic paths : the isobar P=0 (with varying temperature) and the isotherm T=2000K for NS2 (with varying pressure). For convenience this length is represented in Fig. \ref{eyring_loglog} as function of the calculated melt viscosity. Along each path (isobar or isotherm) the arrow indicates the direction of the increasing thermodynamic parameter (T or P). Notice that in the representation $\lambda(\eta)$ (Fig. \ref{eyring_loglog}), both curves exhibit a maximum value (22\AA) for both the isotherm and the isobar which expresses a non-trivial behavior of $\eta$ with diffusivity. Clearly, for the two melt compositions and whatever the thermodynamic path, $\lambda$ decreases when the melt viscosity increases. A simple extrapolation suggests that the Eyring relation should be verified only when the viscosity is sufficiently high (for $\eta>$ 100 Poise),i.e. when $\lambda$ becomes of the order of $\simeq$ 5\AA.. This conclusion is in agreement with the viscosity data discussed above. On the other hand, when the viscosity is becoming very low and the diffusion coefficients of network forming ions high ($>$ 10$^{-9}$~m$^2$/s), the length $\lambda$ matches better the values predicted by the Stokes-Einstein equation, $\lambda$ = 2$\pi$d, where d is the diameter of the diffusing molecule ($\simeq$~ 3\AA\ for oxygen). As the Stokes-Einstein relation works well with molecular liquids (Li and Chang, 1955) it is not too surprising that it also leads to reasonable values for silicate liquids at very high temperature (e.g. in approaching the critical point) where the melt is essentially depolymerized and shares some structural similarities with regular liquids. In summary, the Eyring relation is not adapted to the description of the transport properties of silicate melts at superliquidus temperatures. Alternative models have been proposed in the literature (see Mungall, 2002) but a quantitative theory is still lacking.

\subsection{Structural correlations}

\begin{figure}
\includegraphics*[width=0.9\linewidth, keepaspectratio=true, draft=\ddst]{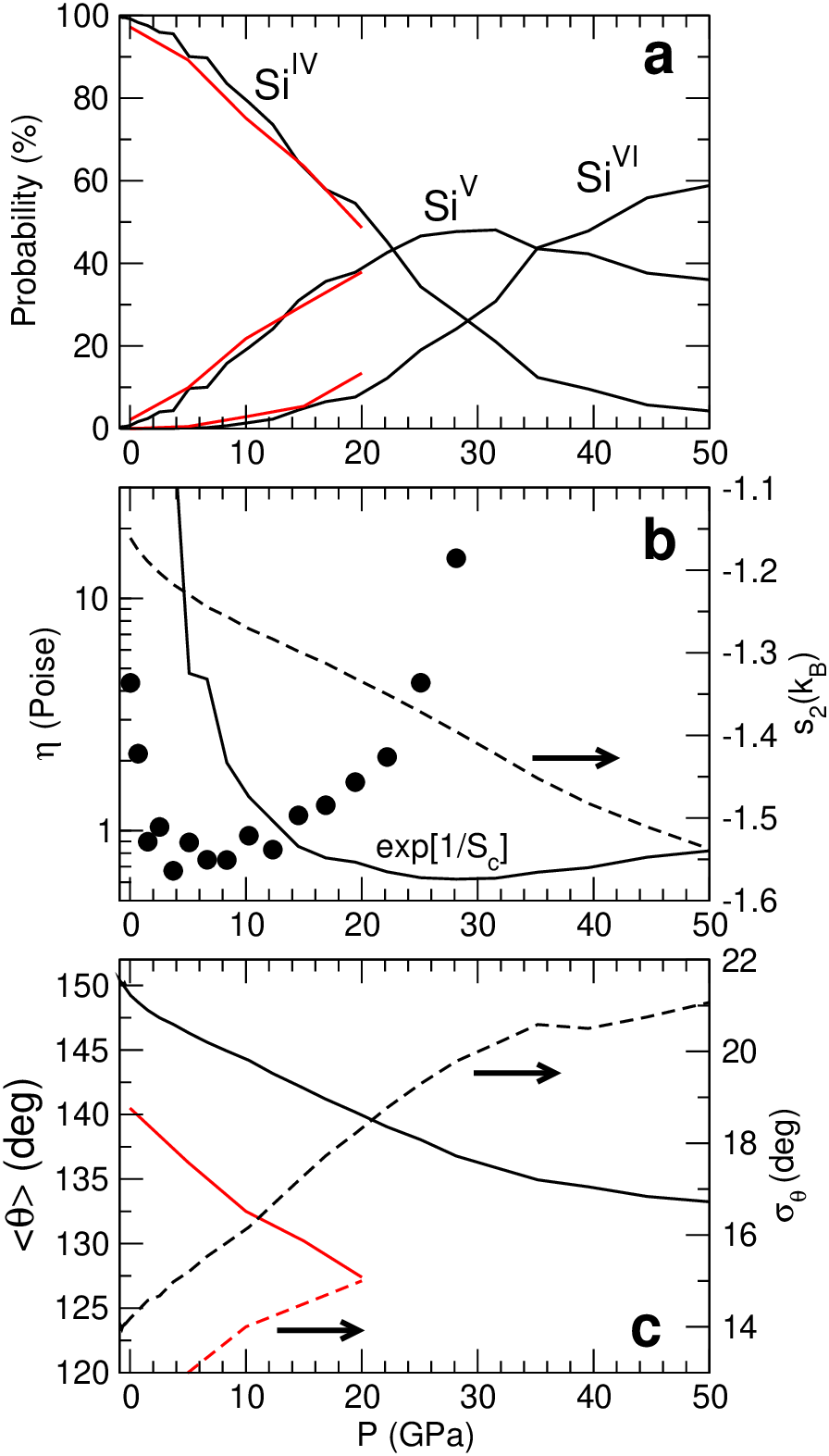}
\caption{\label{entropy} (Color online) a) Fraction of 4-, 5- and 6-fold silicon atoms as a function of pressure in the NS2(black curves) and MORB liquid (red curves). b) Simulated viscosity of the NS2 liquid at T=2000 K (filled black circles, same as Fig. \ref{visco_T}), compared to an Adam-Gibbs viscosity computed from a structure-related configurational entropy S$_c$. Right axis: Entropy s$_2$ (broken line) computed from equ. (\ref{s2}). c) Mean angle around the bridging oxygen (BO) atom and corresponding second moment (standard deviation) of the bond angle distribution (right axis).
}
\end{figure}

As mentioned before, there have been many attempts to correlate structural features with the viscosity anomalies in silicate liquids (Waff, 1975; Woodcock and Angell, 1976; Mysen et al., 1980, Suzuki et al., 2002). From the liquid structures obtained for the two melt compositions (NS2 and MORB), we have computed the fraction of silicon atoms that are 4-, 5- and 6-fold coordinated. Results are displayed in Fig. \ref{entropy}a. Here one sees that the melt structure evolves from roughly 100\% 4-coordinated Si$^{IV}$ to a growing proportion of Si$^V$ and Si$^{VI}$, Si$^V$ increasing substantially for $P $ larger than 10~GPa and going through a maximum found at about 25 GPa in NS2 (and above this pressure for MORB) whereas Si$^{VI}$ is becoming significant only above 15GPa.
\par
From the probability distribution \{p$_{i}$\} of Si$^i$ species (i=IV,V,VI), one can compute a configurational entropy according to:
\begin{eqnarray}
S_c=-\sum_ip_i\ln p_i,
\end{eqnarray}
\par
which maximizes when the population of S$^V$ is maximum ($\simeq$30~GPa).
This configurational entropy is represented in Fig. \ref{entropy}b as $\exp$[1/S$_c$]. In fact, if one assumes that the Adam-Gibbs relation for viscosity of the form: $\eta=\eta_0\exp[A/TS_c]$ is valid, one may expect from a simple inspection of the relationship that a minimum in $\eta$ is correlated with a maximum in $S_c$, as also suggested by Bottinga and 
Richet (1995). However, for NS2 a quick look at figure \ref{entropy}b clearly shows that $\eta$(P) and $\exp$[1/S$_c$] are not correlated as the location of the maximum in S$_c$ does not coincide with the pressure range where a viscosity minimum is observed. In the case of MORB, the gradual increase of the viscosity with pressure could be correlated with that of $S_c$. So it is hard to firmly conclude that a structural reorganization through coordination changes of Si is directly responsible of the pressure evolution of the viscosity. To be complete, notice that Goel et al. (2011) have recently shown by MD simulations that the diffusion coefficient of Si atoms in silicate melts along the MgO-SiO$_2$ join, correlates well with the excess entropy when the latter one is evaluated from the knowledge of the pair distribution functions. But it is unclear if this correlation also holds with the viscosity as the relationship linking diffusivity and viscosity is far to be obvious (see the above discussion). 
\par
We have therefore calculated an excess entropy s$_2$ in using the expression of the two-body term (Baranyai and Evans, 1989), which writes:
\begin{eqnarray}
\label{s2}
s_2 = -{\frac {k_B\rho}{2}}\sum_i\sum_j x_ix_j\int \biggr[g_{ij}\ln g_{ij}-(g_{ij}-1)\biggr]d{\bf r}   
\end{eqnarray}
$\rho$ being the density, and $x_i$ the concentration of species $i$.
Results of this calculation are shown in Fig. \ref{entropy}b (right axis). s$_2$ is found to decrease continuously with pressure and a corresponding Adam-Gibbs $\eta$ does not show a viscosity minimum. This behavior is at variance with that obtained by Goel et al. (2011) in their MD simulation study of MgO-2SiO$_2$ (a system related to NS2) where S$_2$ is exhibiting an entropy maximum around 10~GPa, very well correlated with a diffusivity maximum, as also found by Jabes et al. (2010) for liquid SiO$_2$, GeO$_2$, BeFe$_2$ and H$_2$O. So our search for an eventual relationship between viscosity and configurational entropy at the single temperature 2000~K is inconclusive and certainly deserves a further scrutiny.
An alternative way to understand the trends of the viscosity with pressure may arise from the evolution of the T-O-T intertetrahedral bond angle distribution (BAD) as it is well known that bond angles are much more pressure sensitive than bonds themselves (Micoulaut, 2004). 
Figure \ref{entropy}c shows the evolution of the average value of the BAD between a bridging oxygen (BO) and its two SiO$_{4/2}$ tetrahedra, together with the second moment (standard deviation) of the BAD. The latter one quantifies the angular excursion around a mean value and it is expected to be pressure sensitive, as highlighted in the case of densified GeSe$_2$ (Bauchy et al., 2011). When the average value of the BAD decreases steadily in NS2 melt from 150$^o$ to 140$^o$ between P=0 and 20 GPa, it varies between 140$^o$ and 128$^o$ in MORB over the same pressure range. In the same way the angular excursion increases (with pressure) equally well in the NS2 and in MORB melt as shown by the standard deviation $\sigma_\theta$ evolution with pressure. So, here again, it is not clear that viscosity and diffusion anomalies (e.g. in NS2) induced by the pressure can be univoquely associated with a particular structural rearrangement even if there is no doubt that structural changes are taking place in the silicate melt under investigation. 

\section{Summary and conclusion}
In summary, we have used Molecular Dynamics simulations to study the dynamics of two systems of importance in materials science and geology : the sodium disilicate (NS2) and a basaltic (MORB) liquid. Results show that the diffusion constant computed from a two-body Born-Majer type potential at ambient conditions exhibits a remarkable agreement with experimental data for all species (Na, Si, O) in the NS2 liquid. This contrasts with the level of agreement obtained for the MORB system which shows a deviation by at least one order of magnitude when compared to experimental findings. The same situation is nearly reproduced in the comparison of calculated viscosity using the Green-Kubo formalism (based on the stress auto-correlation function) with experimental data from Bockris et al; (1955) for NS2, and Villeneuve et al. (2008) for MORB.  
\par
We have then studied the behavior of both the diffusion constant and the viscosity with changing pressure in the two liquids with temperature close to 2000~K. In the NS2 system, a diffusivity maximum for the network forming species (Si,O) is obtained at around 15~GPa, a result which parallels similar findings for other densified tetrahedral liquids such as water, germania, or silica. The sodium cation is found to display three distinctive r\'egimes for diffusion, the one corresponding to lower densities (negative pressure) being identified with a stretched melt (Bauchy and Micoulaut, 2011). These features are not found in the more depolymerized MORB liquid which shows for Si and O diffusion a steady behavior between 0 and 5~GPa, followed by a continuous decrease applied pressure. Interestingly, a pressure window is found in the NS2 liquid with a minimum at around 5~GPa that does not seem to be correlated with the maximum in the obtained diffusivity. Again, this contrasts with the steadily increase of the MORB liquid viscosity with pressure. 
\par
Having in hand both the computed diffusion and the viscosity, we have investigated the validity of the Eyring equation which relates both quantities and involves also a jump distance $\lambda$ usually of the order of the average oxygen-oxygen bond distance (2.8~\AA), as exemplified by numerous experimental studies. This investigation has been performed along two thermodynamic paths: at fixed pressure (P=0) and changing temperature, and at fixed temperature (T$\simeq$ 2000 K) and changing pressure. We have shown that $\lambda$ calculated from the simulated diffusion and viscosity was indeed of the order of a few Angstroems, but only in the high viscosity r\'egime (here 100 Poise). It underscores the fact that the Eyring relationship should hold only for deeply supercooled liquids as it fails for liquidus temperatures.
\par
Finally, we have attempted to relate the obtained anomalies in viscosity with structural changes of the melt. With growing pressure, the population of higher coordinated silicon (Si$^V$, Si$^{VI}$) increases in a similar fashion in the NS2 and MORB liquid which rules out the possibility of a direct relationship between e.g. coordination increase and viscosity minimum. Moreover, in the NS2 liquid configurational entropies have been calculated and their behavior with pressure does not follow at all the trend obtained for the viscosity. More subtle structural and thermodynamic changes are clearly at play and one does not recover for NS2 the reported correlation between configurational entropy and diffusivity maxima found by Goel et al. (2011) in the MgO-SiO$_2$ system. 
\par 
Our attempts to find out a quantitative link between pressure-induced structural rearrangement and viscosity anomalies is rather deceptive. Although some correlations between diffusivity, viscosity and excess entropy certainly do exist and are sometimes emphasized in the literature, their effectiveness depends strongly on the melt composition, a fact which makes risky a clear conclusion. On the other hand, a recent theoretical analysis by Schmelzer et al. (2005) seems to be promising. In this study the pressure dependence of the viscosity is analyzed from a thermodynamical point of view. The working equation expresses the derivative of the viscosity with respect to the pressure under the following form,                                                                                                     \begin{eqnarray}
\label{fin}
\biggl({\frac {d\eta}{dP}}\biggl)_T=-{\frac {\kappa_T}{\alpha_T}}\biggl({\frac {\partial\eta}{\partial T}}\biggr)_P
+\biggl({\frac {\partial\eta}{\partial \xi}}\biggr)_T\biggr({\frac {d\xi}{dP}}\biggr)_T ,
\end{eqnarray}

where $\xi$ is an order parameter (e.g. the degree of polymerization given by BO/(BO+NBO) where BO and NBO are the numbers of bridging and nonbridging oxygen, respectively), $\kappa_T$ is the compressibility and $\alpha_T$ the thermal expansion coefficient. If the order parameter is invariant with pressure (so d$\xi$/dP=0) then d$\eta$/dP is governed by the first term in eq. (\ref{fin}) which expresses a free volume effect ($\eta$ only varies with the total volume of the melt). But as $\kappa_T$ is negative in a liquid at equilibrium, and $\partial\eta\partial T$ is positive (the viscosity decreases when the temperature increases), the variation of $\eta$ with P will be positive or negative according to the sign of the thermal expansion coefficient, $\alpha_T$. If the latter one is negative then the viscosity will tend to decrease in increasing pressure and if $\alpha_T$ becomes positive after a further compression then the viscosity will go to a minimum value before to increase. As shown by Schmelzer et al. (2005) this behavior is encountered with the viscosity of liquid water. However the first term in eq. (\ref{fin}) is not sufficient to reproduce all the magnitude of the viscosity minimum and the contribution of structural changes induced by the pressure has to be accounted for (via the second term in eq. (\ref{fin})). In the case of silicate melts it is well documented that the viscosity increases when the degree of polymerization of the melt increases (so d$\eta$/d$\xi>$0) wheras for a given melt its degree of polymerization decreases when pressure increases (so d$\xi$/dP$<$0). Therefore the second term of eq. (\ref{fin}) is generally negative. For NS2 and MORB we have checked that $\alpha_T$ is positive at liquid temperature and over a large pressure range which implies that the first term in eq. (\ref{fin}) is positive for these two compositions. As our MD results for NS2 lead to a decrease of the viscosity between 0 and $\simeq$4GPa, this suggests that the second term in eq. (\ref{fin}) overbalances the first term. For MORB the viscosity is found virtually constant in the 0-4~GPa range so the two terms nearly compensate each other in eq. (\ref{fin}). At high pressures (e.g. above 10 or 20~GPa) the two melts are more and more depolymerized and d$\xi$/dP becomes small because the degree of depolymerization does not evolve very much with the pressure due to melt densification. So the free volume effect (first term) is the dominant contribution to eq. (\ref{fin}) at high compression rates and the viscosity of the two melts increases steadily in the high pressure range. In summary it appears that eq. (\ref{fin}) is very useful to decipher the complex behavior exhibited by the viscosity of silicate melts as function of pressure. To be more quantitative and predictive it remains to make the link between the structural evolution of the melt and the variation of the order parameter with the pressure ($\xi$(P)). This will be done in a future publication.

\section*{Acknowledgments}
The authors thank G. Henderson, D.R. Neuville, P. Richet for useful discussions. 

$$ $$
Adjaoud O., Steinle-Neumann, Jahn S. (2008). Mg2SiO4 liquid under high pressure from molecular dynamics. Chem. Geol. 256, 185-192.
\par
Adjaoud O., Steinle-Neumann, Jahn S. (2011). Transport properties of Mg2SiO4 liquid at high pressure: Physical state of a magma ocean. Earth and Planet. Sci. Lett. 312, 463-470.
\par
Abramo M. C., Caccamo C., Pizzimenti G. (1992). Structural properties and medium-range order in metasilicate (CaSiO$_3$) glass: A molecular dynamics study. J. Chem. Phys. 96, 9083.
\par
Alejandre J., Tildesley D., Chapela G. A. (1995). Molecular Dynamics simulation of the orthobaric densities and surface tension of water. J. Chem. Phys. 102, 4574.
\par 
Allen M. P., Tildesley D. J. (1987). Computer Simulations of Liquids, Oxford Univ. Press, p. 64.
\par
Allen M. P., Brown D., Masters A. J. (1994). Use of the McQuarrie equation for the computation of shear viscosity via equilibrium molecular dynamics - comment. Phys. Rev. E 49, 2488.
\par
Ando R., Ohtani E., Suzuki A., Rubie D.C., Funakoshi K. (2003). Pressure dependency of viscosities of MORB melts. AGU Fall Meeting 2003, abstract V31D-0958.
\par
Angell C.A. (1995). Formation of Glasses from Liquids and Biopolymers. Science 267, 1924.
\par
Ardia P., Giordano D., Schmidt M.W. (2008). A model for the viscosity of rhyolite as a function of H2O-content and pressure: A calibration based on centrifuge piston cylinder experiments. Geochim. Cosmochim. Acta 72, 6103-6123.
\par
Barrat J.-L., Badro J., Gillet P. (1997). Strong to Fragile transition in a model of liquid silica. Mol. Simul. 20, 17-25.
\par
Bauchy M., Micoulaut M. (2011). From pockets to channels: density controlled diffusion in silicates. Phys. Rev. B 83, 184118.
\par
Bauchy M., Micoulaut M., Celino M., Boero M., Le Roux S., Massobrio C. (2011). Angular rigidity in tetrahedral network glasses with changing composition. Phys. Rev. B 83, 054201.
\par
Behrens H., Schulze F. (2003). Pressure dependence of melt viscosity in the system NaAlSi$_3$O$_8$-CaMgSi$_2$O$_6$. Am. Mineral. 88, 1351.
\par
Bockris J.O'M., Mackenzie J.D., Kitchener J.A. (1955). Viscous flow in silica and binary liquid silicates. Trans. Faraday Soc. 51, 1734-1748.
\par
Boon J. P., Yip S. (1980. Molecular Hydrodynamics. McGraw Hill.
\par
Borodin O., Smith G. B., Kim H. (2009). Viscosity of a Room Temperature Ionic Liquid: Predictions from Nonequilibrium and Equilibrium Molecular Dynamics Simulations. J. Phys. Chem. B 113, 4771.
\par
Bottinga Y., Richet P. (1995). Silicate melts: The "anomalous" pressure dependence of the viscosity. Geochim. Cosmochim. Acta 59, 2725-2731.
\par
Brearley M., Dickinson Jr. J.E., Scarfe C.M. (1986). Pressure dependence of melt viscosities on the join diopside-albite. Geochim. Cosmochim. Acta 50, 2563-2570.
\par
Cherne F. J., Deymier P. A. (1998). Calculation of Viscosity of Liquid Nickel byMolecular Dynamics Methods. Scripta Mater 39, 1613. 
\par
Cormack A. N., Du J., Zeitler T. R. (2003). Sodium ion migration mechanisms in silicate glasses probed by molecular dynamics simulations. J. Non-Cryst. Solids 323, 147.
\par
Del Gaudio P., Behrens H. (2009). An experimental study on the pressure dependence of viscosity in silicate melts. J. Chem. Phys. 131, 044504-1-14.
\par
Dingwell D.B., Webb S.L. (1989). Structural relaxation in silicate melts and non-newtonian melt rheology in geologic processes. Phys. Chem. Minerals 16, 508-516.
\par
Dreyfus C., Micoulaut M. (2012), unpublished
\par
Du J., Cormack A. N. (2004.) The medium range structure of sodium silicate glasses: a molecular dynamics simulation. J. Non-Cryst. Solids 349, 66.
\par
Du J., Cormack A. N. (2005). Molecular Dynamics Simulation of the Structure and Hydroxylation of Silica Glass Surfaces. J. Am. Ceram. Soc. 88, 2532.
\par
Du J., Corrales L. R. (2006). Compositional dependence of the first sharp diffraction peaks in alkali silicate glasses: A molecular dynamics study. J. Non-Cryst. Solids, 352, 3255.
\par
Dunn T. (1982). Oxygen diffusion in three silicate melts along the join diopside-anorthite. Geochim. Cosmochim. Acta 46, 2293-2299.
\par
Dunn T., Scarfe C.M. (1986). Variation of the chemical diffusivity of oxygen and viscosity of an andesite melt with pressure at constant temperature. Chem. Geol. 54, 203-215.
\par
Errington J. R., Debenedetti P. G. (2001). Relationship Between Structural Order and the Anomalies of Liquid Water. Nature 410, 259.
\par
Fernandez G. A., Vrabec J., Hasse H. (2004). A Molecular Simulation Study of Shear and Bulk Viscosity and Thermal Conductivity of Simple Real Fluids. Fluid Phys. Equ. 221, 157.
\par
Giordano D., Dingwell D.B. (2003a). The kinetic fragility of natural silicate melts. J. Phys. Condens. Matter 15, S945-S954.
\par
Giordano D., Dingwell D.B. (2003b). Non-Arrhenian multicomponent melt viscosity: a model. Earth Planet. Sci. Letters 208, 337-349.
\par
Giordano D., Russell J.K., Dingwell D.B. (2008). Viscosity of Magmatic Liquids: A Model. Earth Sci. Planet. Sci. Lett. 271, 123-134. 
\par
Goel G., Lacks D.J., Van Orman J.A. (2011). Transport coefficients in silicate melts from structural data via a structure-thermodynamics-dynamics relationship; Phys. Rev. E 84, 051506-1-5.
\par
Ghosh D.B., Karki B.B. (2011). Diffusion and viscosity of Mg2SiO4 liquid at high pressure from first-principles simulations. Geochim. Cosmochim. Acta 75, 4591-4600.
\par
Green M. S. (1954). Markoff random process and the statistical mechanics of time-dependent phenomena 2: irreversible processes in fluids. J. Chem. Phys. 22, 398. 
\par
Guillot B., Sator N. (2007a). A computer simulation study of natural silicate melts. Part I: Low pressure properties. Geochim. Cosmochim. Acta 71, 1249-1265.
\par
Guillot B., Sator N. (2007b). A computer simulation study of natural silicate melts. Part II: High pressure properties. Geochim. Cosmochim. Acta 71, 4538-4556.
\par
Guissani Y., Guillot B. (1994). A numerical investigation of the liquid-vapor coexistence curve of silica. J. Chem. Phys. 104, 7633-7644.
\par
Gupta Y. P., King T. B. (1967). Self-diffusion of sodium in sodium silicate liquids. Trans. Metall. Soc. AIME 239, 1701.
\par
Gupta P. K. (1987). Negative pressure dependence of viscosity. J. Am. Ceram. Soc. 70, C152.
\par
Hemmati M., Angell C.A. (1997) IR absorption of silicate glasses studied by ion dynamics computer simulation. I. IR spectra of SiO$_2$ glass in the rigid ion model approximation. J. Non-Cryst. Solids, 217 236.
\par
Henderson P., Nolan J., Cunningham G.C., Lowry R.K. (1985). Structural controls and mechanisms of diffusion in natural silicate melts. Contrib. Mineral. Petrol. 89, 263-272.
\par
Horbach J., Kob W. (1999). Static and dynamic properties of a viscous silica melt. Phys. Rev. B 60, 3169-3181.
\par
Horbach J., Kob W., Binder K. (2001) Structural and dynamical properties of sodium silicate melts: An investigation by molecular dynamics computer simulation. Chem Geol. 174, 87-101.
\par
Ispas S., Benoit M., Jund P., Jullien R. (2002). Structural properties of glassy and liquid sodium tetrasilicate: comparison between ab initio and classical molecular dynamics simulations. J. Non-Cryst. Solids, 307, 946.
\par
Jabes Shadrack B., Agarwal M., Chakravarty C. (2010). Tetrahedral order, pair correlation entropy, and waterlike liquid state anomalies: Comparison of GeO$_2$ with BeF$_2$, SiO$_2$, and H$_2$O. J. Chem. Phys. 132, 234507.
\par
Johnson J. R., Bristow R. H., Blau H. H. (1951). Diffusion of Ions in Some Simple Glasses. J. Am. Ceram. Soc. 34, 165.
\par
Karki B.B., Stixrude L. (2010). First-principles study of enhancement of transport properties of silica melt by water. Phys. Rev. Lett. 104, 215901-1-4.
\par
Karki B.B., Stixrude L. (2010). Viscosity of MgSiO3 liquid at Earth's mantle conditions: Implications for an early magma ocean. Science 328, 740-742.
\par
Karki B.B., Bohara B., Stixrude L. (2011). First-principle study of diffusion and viscosity of anorthite (CaAl$_2$Si$_2$O$_8$) liquid at high pressure. Am. Mineral. 96, 744-751.
\par
Knoche R., Dingwell D. B., Seifert S. A., Webb S. L. (1994). Non-linear properties of supercooled liquids in the system Na$_2$O-SiO$_2$. Chem. Geol. 116, 1.
\par
Kramer G. J., de Man A. J. M., van Santen R. A. (1991). Zeolites versus aluminosilicate clusters: the validity of a local description. J. Am. Chem. Soc. 113, 6435.
\par
Kubo R. (1957). Statistical-mechanical theory of irreversible processes 1: General theory and simple applications to magnetic and conduction problems. J. Phys. Soc. Jpn. 12, 570.
\par
Kushiro I. (1976). Changes in viscosity and structure of melt of NaAlSi$_2$O$_6$ composition at high pressures. J. Geophys. Res. 81, 6347-6350.
\par
Kushiro I., Yoder Jr. H.S., Mysen B.O. (1976). Viscosities of basalt and andesite melts at high pressures. J. Geophys. Res. 81, 6351-6356.
\par
Kushiro I. (1978). Viscosity and structural changes of albite (NaAlSi$_3$O$_8$) melt at high pressures. Earth Planet. Sci. Letters 41, 87-90.
\par
Kushiro I. (1986). Viscosity of partial melts in the upper mantle. J. Geophys. Res. 91, 9343-9350.
\par
Lacks D.J., Rear D.B., Van Orman, J.A. (2007). Molecular dynamics investigation of viscosity, chemical diffusivities and partial molar volumes of liquids along the MgO-SiO$_2$ join as functions of pressure. Geochim. Cosmochim. Acta 71, 1312-1323.
\par
Lee S. H. (2007). Molecular Dynamics Simulation Study of the Transport Properties of Liquid Argon:
The Green-Kubo Formula Revisited. Bull. Korean Chem. Soc. 28, 1371.
\par
Lees A. W., Edwards S. F. (1972). Computer study of transport processes under extreme conditions. 
J. Phys. Chem. Solid State Phys. 5, 1921.
\par
Lesher C.E., Hervig R.L., Tinker D. (1996). Self diffusion of network formers (silicon and oxygen) in naturally occurring basaltic liquid. Geochim. Cosmochim. Acta 60, 405-413.
\par
Li J.C.M., Chang P. (1955). Self-diffusion coefficient and viscosity in liquids. J. Chem. Phys. 23, 518-520.
\par
Liebske C., Schmickler B., Terasaki H., Poe B.T., Suzuki A., Funakoshi K.-I., Ando R., Rubie D.C. (2005). Viscosity of peridotite liquid up to 13 GPa: Implications for magma ocean viscosities. Earth Planet. Sci. Lett. 240, 589-604.
\par
Melosh H.J. (2007). A hydrocode equation of state for SiO2. Meteor. Planet. Science 42, 2079-2098.
\par
Micoulaut M., Phillips J.C. (2003). Ring statistics and rigidity transitions in network glasses. Phys. Rev. B 67, 104204(1-9). 
\par
Micoulaut M. (2004). Structure of densified amorphous germanium dioxide. J. Phys. Cond. Matt. 16, L131.
\par
Micoulaut M., Guissani Y., Guillot B. (2006). Simulated structural and thermal properties of glassy and liquid germania. Phys. Rev. E 73, 031504.
\par
Micoulaut M. (2010). Linking rigidity with enthalpic changes at the glass transition and the fragility of glass-forming liquids: insight from a simple oscillator model. J. Phys. Cond. Matt. 22, 285101.
\par
Mori S., Ohtani E., Suzuki A. (2000). Viscosity of the albite melt to 7 GPa at 2000 K. Earth Planet. Sci. Letters 175, 87-92.
\par
Mungall J.E. (2002). Empirical models relating viscosity and tracer diffusion in magmatic silicate melts. Geochim. Cosmochim. Acta 66, 125-143 (2002).
\par
Mysen B.O., Virgo D., Scarfe C.M. (1980). Relations between the anionic structure and viscosity of silicate melts- a Raman spectroscopic study. Am. Mineral. 65, 690-710.
\par
Mysen B.O., Richet P. (2005). Silicate glasses and melts: properties and structure, Elsevier.
\par
Negodaev G. D., Ivanov I. A., Evstop'ev K. K. (1972). Elektrokhim. 8, 234.
\par
Ogawa H., Shiraishi Y., Kawamura K., Yokokawa T. (1990). Molecular dynamics study on the shear viscosity of molten Na$_2$O-2SiO$_2$. J. Non Crystal. Sol. 119, 151-158.
\par
Oishi Y., Terai R, Ueda H. (1975). Oxygen diffusion in liquid silicates and relation to their viscosity. In Mass Transport Phenomena in Ceramics (eds. A.R. Cooper and A.H. Heuer) Plenum Press (N.Y.), 297-310.
\par
Pathmanathan K., Johari G.P. (1990). Temperature-dependence of molecular relaxation rates and of viscosity of glass-forming liquids. Phil. Mag. B 62, 225.
\par
Poe B.T., McMillan P.F., Rubie D.C., Chakraborty S., Yarger J., Diefenbacher J. (1997). Silicon and oxygen self-diffusivities in silicate liquids measured to 15 Gigapascals and 2800 Kelvin. Science 276, 1245-1248.
\par
Reid J.E., Poe B.T., Rubie D.C., Zotov N., Wiedenbeck M. (2001). The self-diffusion of silicon and oxygen in diopside (CaMgSi$_2$O$_6$) liquid up to 15 GPa. Chem. Geol. 174, 77-86.
\par
Reid J.E., Suzuki A., Funakoshi K.-I., Terasaki H., Poe B.T., Rubie D.C., Ohtani E. (2003). The viscosity of CaMgSi$_2$O$_6$ liquid at pressures up to 13 GPa. Phys. Earth. Planet Int. 139, 45-54.
\par
Russell J.K., Giordano D., Dingwell D.B. (2003). High-temperature limits on viscosity of non-Arrhenian silicate melts. Am. Mineral. 88, 1390-1394.
\par
Scarfe C.M. (1973). Viscosity of basic magmas at varying pressure. Nature 241, 101-102.
\par
Scarfe C.M., Mysen B.O., Virgo D. (1987). Pressure dependence of the viscosity of silicate melts. In {\em Magmatic processes: Physico-chemical processes} (Ed. B.O. Mysen, Spec. Publ. 1, 59-67, Geochem. Soc.)
\par
Schaefer L., Fegley Jr. B. (2004). A thermodynamic model of high temperature lava vaporization on Io. Icarus 169, 216-241.
\par
Schlenz H., Kirfel A., Schulmeister K., Wartner N., Mader W., Raberg W., Wandelt W., Oligschleger C., Bender S., Franke R., Hormes J., Hoffbauer W., Lansmann V., Jansen M., Zotov N., Marian C., Putz H., Neuefeind J. (2002). Structure analyses of Ba-silicate glasses. J. Non-Cryst. Solids 297, 37.
\par
Schaefer L., Fegley Jr. B. (2004). A thermodynamic model of high temperature lava vaporization on Io. Icarus 169, 216-241.
\par
Schmelzer J.W.P., Zanotto E.D., Fokin V.M. (2005). Pressure dependence of viscosity. J. Chem. Phys. 122, 074511-1-11.
\par
Sharma S. K., Virgo D., Kushiro I. (1979). Relationship between density, viscosity and structure of GeO$_2$ melts at low and high pressures. J. Non-Cryst. Solids 33, 235.
\par
Shell M. S., Debenedetti P. G., Panagiotopoulos A. Z. (2002). Molecular Structural Order and Anomalies in Liquid Silica. Phys. Rev. E 66, 011202.
\par
Shimizu N., Kushiro I. (1984). Diffusivity of oxygen in jadeite and diopside melts at high pressures. Geochim. Cosmochim. Acta 48, 1295-1303.
\par
Shimizu N., Kushiro I. (1991). The mobility of Mg, Ca and Si in diopside-jadeite liquids at high pressures. In Physical Chemistry of Magmas. Advances in Physical Geochemistry, Vol.9 (eds. L.L. Perchuk and I. Kushiro), pp. 192-212. Springer-Verlag, New York.
\par
Smith W., Forester T. (1996). DL\_POLY\_2.0: a general-purpose parallel molecular dynamics simulation package. J. Mol. Graph. 14, 136-141.
\par
Spera F.J., Ghiorso M.S., Nevins D. (2011). Structure, thermodynamic and transport properties of liquid MgSiO$_3$: Comparison of molecular models and laboratory results. Geochim. Cosmochim. Acta 75, 1272-1296.
\par
Suzuki A., Ohtani E., Funakoshi K., Terasaki H. (2002). Viscosity of albite melt at high pressure and high temperature. Phys. Chem. Minerals 29, 159-165.
\par
Suzuki A., Ohtani E., Terasaki H., Funakoshi K. (2005). Viscosity of silicate melts in CaMgSi$_2$O$_6$-NaAlSi$_2$O$_6$ system at high pressure. Phys. Chem. Minerals. 32, 140-145.
\par
Suzuki A., Ohtani E., Terasaki H., Nishida K., Hayashi H., Sakamaki T., Shibazaki Y., Kikegawa T. (2011). Pressure and temperature dependence of the viscosity of a NaAlSi$_2$O$_6$. Phys. Chem. Minerals 38, 59-64.
\par
Teter J. (2003). unpublished results.
\par
Tinker T., Lesher C. E. (2001). Self diffusion of Si and O in dacitic liquid at high pressures. Am. Mineral. 86, 1.
\par
Tinker D., Lesher C.E., Hutcheon I.D. (2003). Self-diffusion of Si and O in diopside-anorthite melt at high pressures. Geochim. Cosmochim. Acta 67, 133-142.
\par
Tinker D., Lesher C.E., Baxter G.M., Uchida T., Wang Y. (2004). High-pressure viscometry of polymerized silicate melts and limitations of the Eyring equation. Am. Mineral. 89, 1701-1708.
\par
Toplis M.J., Dingwell D.B. (2004). Shear viscosities of CaO-Al$_2$O$_3$-SiO$_2$ and MgO-Al$_2$O$_3$-SiO$_2$ liquids: Implications for the structural role of aluminium and the degree of polymerisation of synthetic and natural aluminosilicate melts. Geochim. Cosmochim. Acta 68, 5169-5188.
\par
Trachenko K., Dove M.T., Brazhkin V.V., El'kin F.S. (2004) Network rigidity and properties of SiO2 and GeO2 glasses under pressure, Phys. Rev. Lett. 93, 135502.
\par
Truhlarova M., Veprek O. (1970). Dissolving of vitreous SiO$_2$ in various types of glass
melts under conditions of free convection. Silikaty 14, 1-5. 
\par
Vaills Y., Luspin Y., Hauret G. (2001) Annealing effects in SiO$_2$–Na$_2$O glasses investigated by Brillouin scattering. J. Non-Cryst. Solids 286, 224.
\par
Vaills Y., Qu T., Micoulaut M., Chaimbault F., Boolchand B. (2005). Direct evidence of rigidity loss and self-organization in silicates. J. Phys. Cond. Matt. 17, 4889-4896.
\par
Vessal B., Greaves G.N., Marten P. T., Chadwick A. V., Mole R., Houde–Walter S. (1992). Cation microsegregation and ionic mobility in mixed alkali glasses. Nature 356, 504.
\par
Villeneuve N., Neuville D.R., Boivin P., Bach\`elery P., Richet P. (2008). Magma crystallization and viscosity: A study of molten basalts from the Piton de la Fournaise volcano (La Réunion island). Chem. Geol. 256, 242-251.
\par
Vuilleumier R., Sator N., Guillot B. (2009). Computer modeling of natural silicate melts: What can we learn from ab initio simulations. Geochim. Cosmochim. Acta 73, 6313-6339.
\par
Waff H. S. (1975). Pressure-induced coordination changes in magmatic liquids. Geophys. Res. Lett. 2, 193.
\par
Wang L.M., Angell C.A., Richert R. (2006). Fragility and thermodynamics in nonpolymeric glass-forming liquids. J. Chem. Phys. 125, 074505.
\par
Woddcock L.V., Angell C.A., Cheeseman P. (1976). Molecular dynamics studies of the vitreous state: Simple ionic systems and silica. J. Chem. Phys. 65, 1565.
\par
Wondraczek L., Behrens H., Yue Y., Deubener J., Scherer G.W. (2007). Relaxation and glass transition in an isostatically compressed diopside glass. J. Am. Ceram. Soc. 90, 1556.
\par
Yinnon H., Cooper A.R. Jr. (1980). Oxygen diffusion in multicomponent glass forming silicates. Phys. Chem. Glasses 21, 204-211.
\par
Yuan X., Cormack A.N. (2001). Local structures of MD-modeled vitreous silica and sodium silicate glasses. J. Non-Cryst. Solids 283, 69.
\par
Zhang L., Jahanshahi S. (1998). Review and modeling of viscosity of silicate melts: Part I. Viscosity of binary and ternary silicates containing CaO, MgO, and MnO. Metall. Mat. Transactions B 29, 177.
\par
Zhang L., Van Orman J.A., Lacks D.J. (2010). Molecular dynamics investigation of MgO-CaO-SiO$_2$ liquids: Influence of pressure and composition on density and transport properties. Chem. Geol. 275, 50-57.
\par
Zheng Q., Mauro J.C., Ellison A.J., Potuzak M., Yue Y. (2011). Universality of the high-temperature viscosity limit of silicate liquids. Phys. Rev. B 83, 212202-1-4.

\end {document}